\documentclass[twocolumn,prl,superscriptaddress,showpacs]{revtex4}
\usepackage{graphicx}% Include figure files
\usepackage{bm}% bold math

\begin{document}

\title{Spontaneous Magnetization and Electron Momentum Density in 3D 
Quantum
Dots}

\author{R. Saniz}
\affiliation{Departamento de Ciencias Exactas, Universidad Cat\'olica
Boliviana, Casilla \#5381, Cochabamba, Bolivia}

\author{B. Barbiellini}
\affiliation{Physics Department, Northeastern University, Boston,
Massachusetts 02115}

\author{A. B. Denison}
\affiliation{Lawrence Livermore National Laboratory,
7000 East Ave., Livermore, CA 94550-9234}

\author{A. Bansil}
\affiliation{Physics Department, Northeastern University, Boston,
Massachusetts 02115}

\date{\today}


\begin{abstract}

We discuss an exactly solvable model Hamiltonian for describing 
the interacting electron gas in a quantum dot. Results for a 
spherical square well confining potential are presented. The ground 
state is found to exhibit striking oscillations in spin polarization 
with dot radius at a fixed electron density. These oscillations are 
shown to induce characteristic signatures in the momentum density 
of the electron gas, providing a novel route for direct experimental 
observation of the dot magnetization via spectroscopies sensitive to 
the electron momentum density.

\end{abstract}

\maketitle

Quantum dots, which may be thought of as artificial ``atoms'', 
not only provide a nanoscale laboratory for fundamental 
quantum mechanical studies, but also hold the promise of 
novel spintronic and other
applications \cite{ashoori96,wolf01,ciorga02,smet02}.
Spin polarization in nanosystems is therefore drawing
great attention in view of its intrinsic 
importance \cite{tarucha96,zabala98,fujisawa02}, 
and its possible relationship to other phenomena, such as
kinks in the positions of Coulomb blockade peaks
in 2D quantum dots \cite{baranger00}, or the
so-called 0.7 conduction anomaly in quantum point 
contacts \cite{wang98}. In 2D dots, experiments point to 
shell-like filling of levels in accord with
Hund's first rule \cite{tarucha96}, and a recent 
spin density-functional calculation indicates the 
presence of a ground state with high spin in relatively 
large dots \cite{jiang02}. In 3D dots, shell structure of 
energy levels has been reported \cite{bakkers01}, and 
results on the fine structure in the energy
spectra of gold nanoparticles have been interpreted 
in terms of a ground state spin magnetization
larger than 1/2 \cite{davidovic00}. However, 
Ref.~\onlinecite{jacquod00} argues that in finite 
disordered systems fluctuations in off-diagonal 
interaction matrix elements may 
suppress ground state magnetization. It is clear that 
spin polarization of quantum dots is a subject of 
considerable current interest and controversy.

In this work, we develop an exactly solvable many-body model
Hamiltonian of wide applicability for describing 
correlated electrons confined to a quantum dot. The ground state is
clearly found to exhibit spontaneous magnetization and strong 
oscillations in spin polarization as a function of the dot radius. 
In order to identify observable signatures 
of the remarkable aforementioned 
oscillations in polarization, we consider the momentum density
of the electron gas in the dot at some length. The key quantity 
in this connection turns out to be the width $\Delta p$ over which 
the electron momentum density falls to zero around the Fermi cut-off 
momentum. In particular, oscillations in magnetization
are shown to induce similar oscillations in $\Delta p$, indicating 
that this effect would be amenable to direct 
experimental observation via electron momentum density 
spectroscopies such as angular correlation of positron 
annihilation radiation (ACAR) and Compton scattering. 
Our results also bear more broadly on the applications of quantum
dots in that they point to the possibility of obtaining a system in a
ferromagnetic ground state without 
requiring low temperatures and/or external magnetic fields.

We consider the Hamiltonian
\begin{equation}
\hat H=\sum_{\nu\sigma}\epsilon^0_\nu 
a^\dagger_{\nu\sigma}a_{\nu\sigma}
+{1\over 2}U\sum_{\nu\nu'\sigma}
a^\dagger_{\nu\sigma}a^\dagger_{\nu'\,-\sigma}
a^{\phantom{dagger}}_{\nu'\,-\sigma}
a^{\phantom{\dagger}}_{\nu\sigma}. \label{hamiltonian}
\end{equation}
The first term ($\hat H_0$) describes the noninteracting system. It
corresponds to the spherical square well:
$V(r)=-V_0$, for $r\leq R$, and $V(r)=0$ otherwise, and the associated
one-particle eigenfunctions ${\varphi_\nu}$ with eigenvalues
$\epsilon_\nu^0$; $\nu\equiv nlm$ is a composite index and $\sigma$
labels spin. The second term ($\hat H_I$) describes the
interaction and allows Coulomb repulsion via the parameter $U$ 
only between electrons of opposite spin---electrons
of like spin being kept apart by the Pauli exclusion principle.
In this connection, we have carried out 
extensive numerical estimates, and find that direct
Coulomb matrix elements are larger, on the average, by 
an order of magnitude or
more, than exchange or other terms 
for dot radii up to 30 {\AA}
and electron densities ranging from $r_s=2$ to 5 (Bohr radii) \cite{comment0}.
Spin-orbit effects
here are also expected to be small, since the confining potential is
relatively weak due to the screening of the nuclei by the
core electrons. For these reasons, form (1) should provide a reasonable 
model for the electronic structure of 3D quantum dots.

The standard one-particle Green function for the many electron system 
can be expanded as
${\cal G}_{\sigma\sigma'}({\bf r}{\bf r}',\tau)=\sum_{\nu\nu'}
\varphi^*_\nu({\bf r})
\varphi_{\nu'}({\bf r}')
{\cal G}_{\sigma\sigma'}(\nu\nu',\tau)$, with
${\cal G}_{\sigma\sigma'}(\nu\nu',\tau)=-\langle T_\tau 
a_{\nu\sigma}(\tau)
a_{\nu'\sigma'}^\dagger(0)\rangle$.
It is readily shown that
$\hbar\partial a_\nu/\partial\tau
=-\xi_\nu a_{\nu\sigma}-
U\hat N_{-\sigma}a_{\nu\sigma}$,
where $\xi_\nu=\epsilon_\nu^0-\mu$ ($\mu$ is the chemical 
potential), and
$\hat N_{-\sigma}=\sum_{\nu'} a^\dagger_{\nu'\,-\sigma}
a^{\phantom{\dagger}}_{\nu'\,-\sigma}$.
Then, the equation of motion for the Green function 
${\cal G}_{\sigma\sigma'}(\nu\nu',\tau)$ is
\begin{eqnarray}
-\hbar{\partial\over{\partial\tau}}{\cal 
G}_{\sigma\sigma'}(\nu\nu',\tau)
&=&\hbar\delta_{\sigma\sigma'}\delta_{\nu\nu'}\delta(\tau)
+\xi_\nu{\cal G}_{\sigma\sigma'}(\nu\nu',\tau) \nonumber \\
&-&U\langle T_\tau\hat N_{-\sigma}(\tau)
a^{\phantom{\dagger}}_{\nu\sigma}(\tau)a^\dagger_{\nu'\sigma'}(0)
\rangle.
\end{eqnarray}
Since our Hamiltonian possesses the property that
$[\hat H_I,\hat H_0-\mu\hat N]=0$, Wick's theorem allows us to 
factorize the preceding ground state average, i.e.,
$\langle T_\tau\hat N_{-\sigma}(\tau)
a^{\phantom{\dagger}}_{\nu\sigma}(\tau)a^\dagger_{\nu'\sigma'}(0)
\rangle=-N_{-\sigma}{\cal G}_{\sigma\sigma'}(\nu\nu',\tau)$, where
$N_{-\sigma}=\sum_{\nu'} f_{\nu'\,-\sigma}$, with
$f_{\nu'\,-\sigma}$ denoting the Fermi occupation function.
A Fourier transform in $\tau$
immediately yields the {\sl exact} Green function
\begin{equation}
{\cal G}_{\sigma\sigma'}(\nu\nu',i\omega_n)
=\delta_{\sigma\sigma'}\delta_{\nu\nu'}
{1\over{i\omega_n-\xi_\nu/\hbar-U\,N_{-\sigma}/\hbar}}.\label{sol}
\end{equation}
The self-energy thus is
$\Sigma^*_\sigma=UN_{-\sigma}/\hbar$.
The interacting energy levels, $\epsilon_{\nu\sigma}$,
are given by the solutions of $\omega-\xi_\nu/\hbar-\Sigma^*_\sigma=0$
(frequencies measured with respect to $\mu/\hbar$)
minimizing the ground state energy, 
$E=\sum_{\nu\sigma}(\epsilon^0_\nu+
{1\over 2}\hbar\Sigma^*_\sigma)f_{\nu\sigma}$.
Given the number of particles $N$,
this defines a set of nonlinear equations for the populations,
$N_\uparrow$ and $N_\downarrow$, of
up and down spin states, respectively.
The resulting splitting in energy for states of opposite spin,
$\Delta=U(N_\uparrow-N_\downarrow)$, is {\sl uniform}, i.e., it
does not depend on the quantum numbers $\nu$. 
Note that temperature enters our formulae only through the
Fermi function, $f_{\nu\sigma}$; all results in the remainder 
of this article refer to the zero temperature limit.  

\begin{figure}
\begin{center}
\includegraphics[width=\hsize]{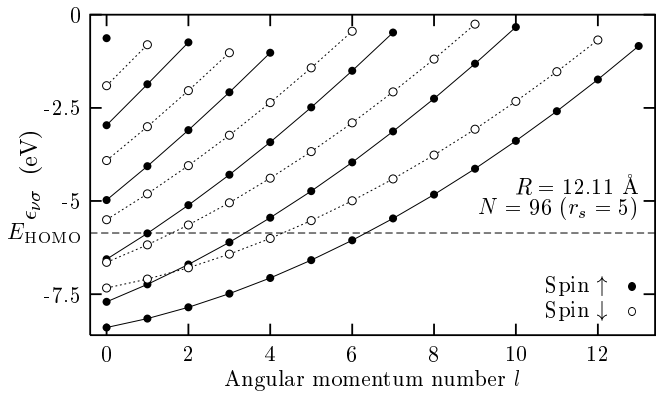}
\end{center}
\caption{Typical energy spectrum $\epsilon_{\nu\sigma}$ 
for the interacting Hamiltonian (1).
Filled (open) circles give majority (minority) spin levels,
those with equal $n$ joined by solid (dotted) lines for clarity. 
Levels possess a degeneracy of $2l+1$.
Horizontal dashed line marks the 
highest occupied molecular orbital energy, $E_{\rm HOMO}$.}
\label{fig1}
\end{figure}

We comment briefly on the nature of the four essential parameters
which describe our model:
$r_s$, the electron density; $R$, the dot radius or, equivalently, the
number $N$ of electrons [$\,N=(R/r_s)^3$];
$V_0$, the well-depth; and, $U$, the Coulomb energy.
The values of $r_s$ and $R$ are related to
the type and size of the systems considered. In this study we use
$r_s=5$, a relatively low density, enabling us to
consider a wide range of
dot radii. $V_0$ may then be obtained reasonably 
in terms of the work function of the system studied and the Fermi 
energy
of the free electron gas with the same $r_s$. 
For $r_s=5$, following
Ref.~\onlinecite{saniz02}, we take $V_0=8.62$ eV.
The choice of the remaining parameter $U$ is somewhat
tricky, mainly because correlations in
quantum dots are not well understood. A handle on $U$
can be obtained by considering the average value $U_{\rm TF}$ of the
Thomas-Fermi (TF) screened Coulomb interaction, 
$\int d{\bf r}d{\bf r}'|\varphi_\nu({\bf r})|^2
v_{\rm TF}(|{\bf r}-{\bf r}'|) |\varphi_\mu({\bf r}')|^2$, averaged 
over all occupied states in the {\sl noninteracting} system,
with $v_{\rm TF}(r)=e^2\exp(-r/l_{\rm TF})/r$, where 
$l_{\rm TF}=\sqrt{r_s}/1.56$ is the TF screening length. 
However, actual screening in real materials is likely 
to be weaker. For example, in noble metals, 
Ref.~\onlinecite{delfatti00}, reports an effective screening length 
$l_{\rm TF}/0.73$. 
Using this value, we find an enhanced Coulomb interaction given roughly 
by $1.75\,U_{\rm TF}$,
which is the value we have used throughout this work.

Fig.~\ref{fig1} illustrates the nature of the energy levels
$\epsilon_{\nu\sigma}$ 
for a 12.11 {\AA} radius dot, for which $N=96$,
and $U= 27.13$ meV ($U_{\rm TF}= 15.50$ meV).
The self-consistent splitting between up and down spin electrons is 
found to
be, $\Delta=1.03$ eV. The highest occupied molecular orbital (HOMO) 
energy
level is, $E_{\rm HOMO}=-5.86$ eV, which may be thought of as 
the dot ``Fermi energy''. We have
$N_\uparrow=67$ and $N_\downarrow=29$, so that the ground state spin
polarization per electron is $\zeta=(N_\uparrow-N_\downarrow)/N=0.4$.

\begin{figure}
\begin{center}
\includegraphics[width=\hsize]{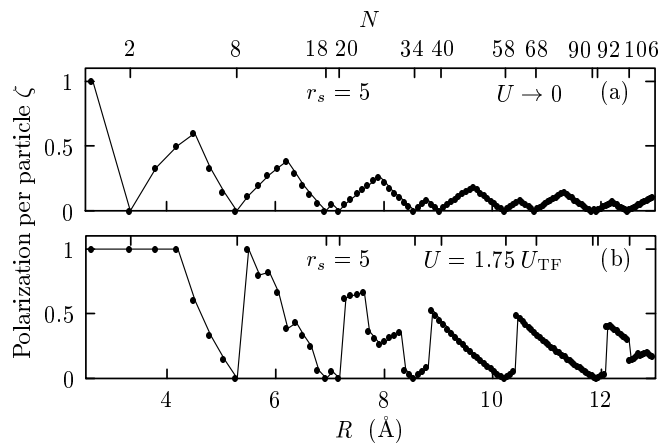}
\end{center}
\caption{Spin polarization per particle,
$\zeta=(N_\uparrow-N_\downarrow)/N$,
as a function of the dot radius $R$. Computed values (filled
circles)
are connected by solid lines to guide the eye. Number of electrons
$N$ is indicated on the top axis. 
(a) Weakly
interacting case, $U\to 0$. (b) Strongly interacting case,
$U=1.75\,U_{TF}$.}
\label{fig2}
\end{figure}

We discuss now $\zeta$ as a function of dot radius with reference to 
Fig.~\ref{fig2}. In the weakly interacting limit 
($U\to 0$), $\Delta\to 0$, still
the degeneracy between up and
down spin in the $nl$ shells is lifted, and as
Fig.~\ref{fig2}(a) clearly shows, 
shells fill in accord with Hund's first rule. Indeed, as the
number of particles increases with dot radius, 
the spin polarization reaches a
peak each time a shell is half filled with up-spin electrons, and
falls to zero when the shell is completed with down-spin electrons.
The sequence of these so-called ``magic numbers'', i.e., $N$ values
for which all occupied shells are completely filled, will of course
differ here from that in
atoms or 2D quantum dots due to differences in the details of the 
underlying spectrum. The magic numbers up to 106 are 
indicated on the upper axis in Fig.~\ref{fig2}(a). 
Turning next to the strongly interacting case, 
Fig.~\ref{fig2}(b) shows large deviations in 
$\zeta$ from a simple Hund's rule filling \cite{comment1}.
This is because $\Delta$ changes with each added electron in order to
minimize the total energy. The first 4 electrons enter the up spin
$1s$ and $1p$ levels, so that $\zeta=1$, while the next 4 enter the
corresponding down spin levels until $\zeta$ vanishes 
for $N=8$. The 9th electron
induces a strong change in $\Delta$, so that the $1s$ and $1p$ down 
spin
levels are pushed above the $1d$ up spin level, causing the former to 
empty
in favor of the latter and the system to be become 
completely polarized again. Similar level
reorderings are involved in the other ``kinks'' seen in 
Fig.~\ref{fig2}(b).

\begin{figure}
\begin{center}
\includegraphics[width=\hsize]{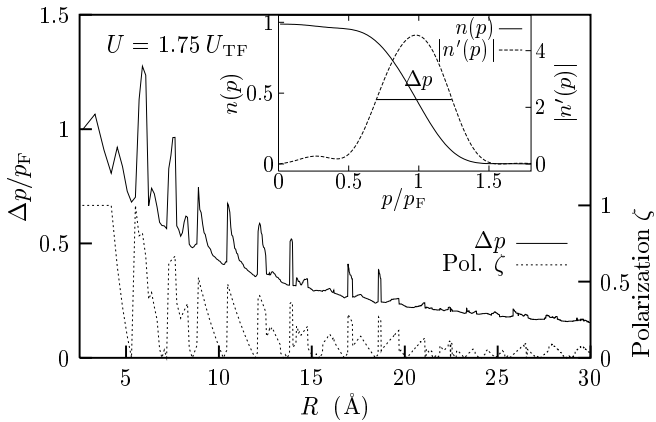}
\end{center}
\caption{
{\it Inset}: Typical electron momentum density, $n(p)$, and
the magnitude of its first derivative,
$|n'(p)|$;
curves are normalized by $V/4\pi^3$, where $V$ is the dot volume.
The position of the peak in $|n'(p)|$ defines the dot ``Fermi
momentum'', while its full-width-at-half-maximum (FWHM)
defines $\Delta p$.
{\it Main graph}:
$\Delta p$ in reduced units of $p_{\rm F}$ (solid line, left scale),
and spin polarization $\zeta$ (dotted, right scale)
{\it vs.} dot radius $R$.}
\label{fig3}
\end{figure}

In order to identify observable signatures of the remarkable changes in 
the polarization depicted in Fig.~\ref{fig2}, we consider the electron
momentum density (EMD), defined by
$n({\bf p})=(2\pi)^{-1}\int_{-\infty}^\infty d\omega\,
f(\omega)A({\bf p},\omega)$,
where $f$ is the Fermi function and $A$ is the spectral function
$A({\bf p},\omega)=-2\,{\rm Im}G^R({\bf p},\omega)$, with
$G^R({\bf p},\omega)={\cal G}({\bf p},i\omega_n\to \omega+i\delta)$.
The inset in Fig. 3 illustrates $n(p)$ and the magnitude of its
first derivative, $|n'(p)|$, for the same parameter 
values as in Fig~\ref{fig1}.
The region of rapid variation in $n(p)$ can be characterized via the
position, $p_{\rm F}$, of the peak in $|n'(p)|$ and the associated
full-width-at-half-maximum (FWHM), $\Delta p$.  In the bulk limit in a 
metallic system $p_{\rm F}$ will tend to the Fermi
momentum, where $n(p)$ will develop a break and $|n'(p)|$ a
$\delta$-function peak, correspondingly. For simplicity, therefore, we 
may
refer to $p_{\rm F}$ loosely as the dot ``Fermi momentum'', even though 
there are no breaks in the EMD in a 
finite system \cite{comment2,comment3}.
Fig.~\ref{fig3} considers $\Delta p$ systematically and shows the 
presence
of a dramatic series of peaks (solid line) spaced more or less 
regularly as 
a function of $R$. A comparison with corresponding changes in $\zeta$ 
(dotted
curve), makes it evident that peaks in $\Delta p$ are well correlated 
with those in $\zeta$, some differences notwithstanding. This 
correlation arises because, at a peak in $\zeta$, the up and down 
spin EMDs become substantially shifted with 
respect to each other, reflecting the polarization
of the system, so that the EMD presents a much larger 
value of $\Delta p$ overall. In short, peaks in $\Delta p$ 
clearly are a signature of peaks in $\zeta$.

\begin{figure}
\begin{center}
\includegraphics[width=\hsize]{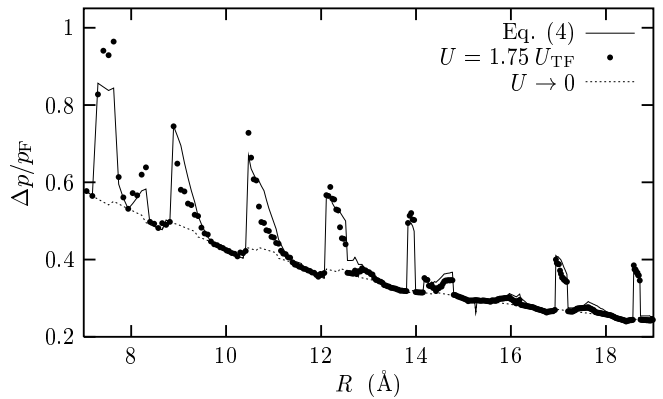}
\end{center}
\caption{
$\Delta p/p_F$ {\it vs.} dot radius $R$,
obtained from the approximate
expression of Eq.~(\ref{fit}) (solid line), compared with
the exact result (bullets). The baseline
corresponding to the noninteracting limit,
$\Delta p_0/p_{\rm F}$, is given (dotted) for reference.}
\label{fig4}
\end{figure}

Motivated by the preceding considerations, 
we have obtained an approximate expression for  
$\Delta p$ as 
\begin{equation}
\Delta p/p_{\rm F}=\Delta p_0/p_{\rm F}
+c\Delta(\zeta-\zeta_0)\,R, \label{fit}
\end{equation}
where $\zeta_0$ is the spin polarization in the weakly 
interacting case ($U\to 0$),
and $c$ is a constant, which depends on various dot parameters,
with a fitted value in the present case of, $c=4.29\times 10^{-2}$ 
eV$^{-1}${\AA}$^{-1}$.
The first term in Eq.~(\ref{fit}) describes the $U\to 0$ limit and 
can be shown to follow the scaling
law $0.93\,{\rm\AA} (r_s/a_0)/R$ \cite{saniz02}.
The second term incorporates the effect of the spin polarization 
induced
through the interaction $U$. 
Fig.~\ref{fig4} shows that Eq.~(\ref{fit}) provides an excellent 
description of the exact $\Delta p$ 
data as a function of $R$. The approximation continues to
be equally good up to $R=30$ \AA, although for clarity
results over a shorter $R$-range are shown in the figure.

We emphasize that the form of our solution 
in Eq.~(\ref{sol}) does not depend explicitly on 
that of the noninteracting Hamiltonian ($\hat H_0$).
Although details of $\zeta$ will vary with 
those of the spectrum of $\hat H_0$, we expect our prediction of  
oscillations in $\zeta$ with dot radius to be generally robust 
to changes in the shape and dimensionality of the 
confining potential \cite{comment5,comment6}.
Thus, we expect our results to be relevant to 
almost one dimensional systems. Indeed, in metallic 
nanowires \cite{comment6b},
Zabala {\it et al.} \cite{zabala98} report a 
series of peaks in the magnetic 
moment per electron as a function of wire radius, correlated
strongly with another observable of the system (the
elongation force of the nanowire). The physics driving these results 
is similar to that in our work \cite{comment7}, despite the fact
that Ref.~\onlinecite{zabala98} considers the stabilized jellium 
model within the framework of the spin-dependent density-functional 
formalism, while we treat correlation effects differently,
in terms of the direct interaction $U$ and neglect 
various off-diagonal exchange and Coulomb
contributions \cite{comment8}.

The relationship between $\zeta$ and $\Delta p$ provides a powerful 
handle for a direct experimental verification of the 
polarization oscillations in quantum dots
predicted in this study via the use of solid state spectroscopies
sensitive to the EMD, namely, angular-correlation of 
positron-annihilation
radiation (ACAR) [positrons have the advantage of being local
probes, see Weber {\it et al.} in Ref.~\onlinecite{comment2}],
inelastic X-ray scattering (IXS) in the deeply
inelastic (Compton) regime and scanning tunneling microscopy 
(STM). We have carried out further computations 
at other $r_s$ values and find that, generally speaking, 
polarization effects become less prominent with decreasing 
$r_s$, as the kinetic energy dominates, as well as with 
increasing $R$, as electrons are less confined and 
$U$ becomes weaker. Nevertheless, 
oscillations in $\Delta p$ are observable even for
$r_s\simeq 3$, particularly for dot sizes $R\alt 10$ \AA.

In summary, we have presented an {\it exactly solvable} model 
Hamiltonian for discussing the properties 
of the interacting electron gas in the confined geometry of 
a quantum dot. Although the results presented in this article 
are based on a confining potential in the shape of a spherical 
square well, computations for other geometries and confining 
potentials would be quite straightforward. 
The ground state is found to exhibit 
spontaneous magnetization and striking oscillations in spin 
polarization $\zeta$ as a function of the dot radius $R$, at the
fixed electron density considered. 
The oscillations in $\zeta$ are shown to induce similar 
oscillations in the width $\Delta p$ around the Fermi cut-off 
in the momentum density, which we refer to for simplicity as the 
dot ``Fermi momentum'',
providing a novel route for direct experimental observation 
of the dot magnetization via spectroscopies sensitive to 
momentum density (specially, positron annihilation and Compton 
scattering). A simple expression for 
$\Delta p$ is discussed, which gives an excellent approximation 
to the exact numerical data as a function of radius. 
We expect the results of the present 
study to be robust to the details 
of the confining potential and to the approximations 
inherent in the treatment of correlations in our model 
Hamiltonian.

%\begin{acknowledgments}
This work was supported by the U. S. Department of Energy
contracts DE-AC03-76SF00098
and W-7405-ENG-48 and the U. S. Department of Energy Office of Basic Energy
Science, Division of Materials Science, and benefitted
from the allocation of supercomputer time at the 
NERSC and the Northeastern University
Advanced Scientific Computation Center (ASCC).
R.S. acknowledges the hospitality and support of 
Northeastern University's Physics Department
during a research visit.
%\end{acknowledgments}


\begin{thebibliography}{99}

\bibitem{ashoori96} R. C. Ashoori, Nature {\bf 379}, 413 (1996);
L. P. Kouwenhoven, C. M. Marcus, P. L. McEuen, S. Tarucha,
R. M. Westervelt, and N. S. Wingreen, in {\it Mesoscopic Electron 
Transport},
NATO Science Series: E, Vol. 345, edited by L. L. Sohn,
L. P. Kouwenhoven, G. Sch\"on, (Kluwer, Dordrecht, 1997);
S. M. Reimann and M. Manninen, Rev. Mod. Phys. {\bf 74},
1283 (2002).

\bibitem{wolf01} S. A. Wolf, D. D. Awschalom, R. A. Buhrman,
J. M. Daughton, S. von Moln\'ar, M. L. Roukes, A. Y. Chtchelkanova,
and D. M. Treger, Science {\bf 294}, 1488 (2001).

\bibitem{ciorga02} M. Ciorga, A. Wensauer, M. Pioro-Ladriere,
M. Korkusinski, J. Kyriakidis, A. S. Sachrajda, and P. Hawrylak,
Phys. Rev. Lett. {\bf 88},
256804 (2002).

\bibitem{smet02} J. H. Smet, R. A. Deutschmann, F. Ertl,
W. Wegscheider, G. Abstreiter, and K. von Klitzing,
Nature {\bf 415}, 281 (2002).

\bibitem{tarucha96} S. Tarucha, D. G. Austing, T. Honda,
R. J. van der Hage, and L. P. Kouwenhoven,
Phys. Rev. Lett. {\bf 77}, 3613 (1996).

\bibitem{zabala98} N. Zabala, M. J. Puska, and R. M. Nieminen,
Phys. Rev. Lett. {\bf 80}, 3336 (1998) and
Phys. Rev. B {\bf 59}, 12652 (1999).

\bibitem{fujisawa02} T. Fujisawa, D. G. Austing, Y. Tokura,
Y. Hirayama, and S. Tarucha, Nature {\bf 419}, 278 (2002).

\bibitem{baranger00} H. U. Baranger, D. Ullmo, and L. I. Glazman,
Phys. Rev. B {\bf 61}, R2425 (2000).

\bibitem{wang98} C.-K. Wang and K.-F. Berggren,
Phys. Rev. B {\bf 57}, 4552 (1998).

\bibitem{jiang02} H. Jiang, H. U. Baranger, and W. Yang,
arXiv:cond-mat/0208146v1.

\bibitem{bakkers01} E. P. A. Bakkers, Z. Hens, A. Zunger,
A. Franceschetti, L. P. Kouwenhoven, L. Gurevich,
and D. Vanmaekelbergh, Nano Lett. {\bf 1}, 551 (2001).

\bibitem{davidovic00} D. Davidovi\'c and M. Tinkham,
Phys. Rev. B {\bf 61}, R16359 (2000).

\bibitem{jacquod00} Ph. Jacquod and A. D. Stone,
Phys. Rev. Lett. {\bf 84}, 3938 (2000).

\bibitem{comment0} For instance, for the dot referred to
in Fig. 1,
the direct bare Coulomb matrix elements average to
$\langle U_{\rm d} \rangle =0.114$ Ry, with a low standard deviation
$\sigma_{\rm d}=0.009$ Ry. Among the next leading terms, in contrast,
the exchange matrix elements average to
$\langle U_{\rm ex} \rangle =0.005$ Ry ($\sigma_{\rm ex}=0.006$ Ry).

\bibitem{saniz02} R. Saniz, B. Barbiellini, and A. Denison,
Phys. Rev. B {\bf 65}, 245310 (2002).

\bibitem{delfatti00} N. Del Fatti, C. Voisin, M. Achermann, S. 
Tzortzakis,
D. Christofilos, and F. Vall\'ee, Phys. Rev. B {\bf 61}, 16956 (2000).
See also I. Nagy, M. Alducin, J. I. Juaristi, and P. M. Echenique,
Phys. Rev. B {\bf 64}, 075101 (2001).

\bibitem{comment1} Note deviations from first Hund's rule filling
in 2D quantum dots
have long been reported. See, e.g., 
Ref.~\onlinecite{tarucha96}.

\bibitem{comment2} One has $\Delta p>0$ also in bulk semiconductors and 
metals along directions in the Brillouin zone where the spectrum 
contains
a band gap (e.g., along [111] in Cu). In these cases the cutoff 
momentum
is related
to the dimension of the Jones or Brillouin zones, respectively.
See M. H. Weber {\em et al.}, Phys. Rev. B {\bf 66}, 041305(R) (2002).

\bibitem{comment3}  The asphericity of the momentum density when there 
are incomplete shells is found to be very small.

\bibitem{comment5} We do not contradict the theorem of Lieb and
Mattis [Phys. Rev. {\bf 125}, 164 (1962)] 
regarding the absence of a magnetic ground state in the 1D case
because our interaction in Eq.~(\ref{hamiltonian})
is spin dependent. 

\bibitem{comment6} Even if the confining potential is deformed,
to observe oscillations in $\zeta$ it will
suffice to have singularities in the noninteracting density of
states and a strong $U$.

\bibitem{comment6b} A nanowire
can be viewed as a very elongated ellipsoid.

\bibitem{comment7} Our $UN$ is close to the Stoner
parameter $I$ of Ref.~\onlinecite{zabala98}, both in meaning and in value.


\bibitem{comment8} These off-diagonal terms are generally expected to 
reduce the ground state polarization.

\end{thebibliography}
\end{document}